# Formation energy profiles of oxygen vacancies at grain boundaries in perovskite-type electroceramics

*Daniel Mutter*[1*], *Cong Tao*[1*], *Daniel F. Urban*[1,2], *Christian Elsässer*[1,2]

[1]Fraunhofer IWM, Wöhlerstraße 11, 79108 Freiburg, Germany
[2]Freiburg Materials Research Center (FMF), University of Freiburg, Stefan-Meier-Straße 21, 79104 Freiburg, Germany

*Corresponding authors:
daniel.mutter@iwm.fraunhofer.de
cong.tao@iwm.fraunhofer.de

## Abstract

Oxygen vacancy formation energies play a major role in the electric field assisted abnormal grain growth of technologically relevant polycrystalline perovskite phases. The underlying effect on the atomic scale is assumed to be a redistribution of cationic and anionic point defects between grain boundaries and the bulk interior regions of the grains due to different defect formation energies in the structurally different regions, accompanied by the formation of space charge zones. Using atomistic calculations based on classical interatomic potentials, we derive and discuss optimized structures of the symmetric tilt grain boundaries $\Sigma 5(210)[001]$ and $\Sigma 5(310)[001]$, and of the asymmetric tilt grain boundary $(430)[001]||(100)[001]$ in the electroceramic perovskite materials $SrTiO_3$, $BaTiO_3$, and $BaZrO_3$. We present profiles of oxygen vacancy formation energies across those GBs and discuss their dependence on composition and grain boundary type.

## 1. Introduction

The material class of electroceramics is widely utilized in functional key components of technological applications, ranging from optoelectronics[1] over ferro-[2,3] and piezoelectric devices[4] to cells for conversion and storage of energy[5]–[7]. A major reason for the remarkable versatility of electroceramics, which are transition metal oxide compounds, is their compositional variability and the implicated variety of possible defects in the structure. Especially oxygen vacancies, which can be present in the materials in large concentration ranges, enable many functionalities, such as electronic or ionic conductivity[8,9].

For the perovskite compound strontium titanate, $SrTiO_3$ (STO), the high diffusivity of oxygen vacancies allows the technologically promising effect of electric-field-assisted sintering, specifically the growth of very large crystallites within the polycrystalline microstructure[10]. For an STO sample attached between oppositely charged electrodes in their experiment, the authors of Ref. [10] stated a concentration gradient of oxygen vacancies. Typically, the Fermi level is located below the charge transition levels of oxygen vacancies, which lie higher in the band gap and above the charge transition levels of acceptor-like cation vacancies. Therefore, under most conditions, the latter are negatively charged while oxygen vacancies are positively charged. In addition, there are concentration differences of oxygen vacancies and cationic vacancies between the grain boundary (GB) cores and the bulk regions of the grains, accompanied by the formation of space charge regions (SCRs) around the GBs[11]. Accordingly, the different defect concentrations at the two electrodes lead to SCRs of different strengths and ranges, which unequally affect the GB mobilities, and hence the grain growth.

Besides the work of Rheinheimer et al.[10] mentioned above, abnormal grain growth behavior in perovskite oxides was reported before in STO[12]–[14] and $BaTiO_3$ (BTO)[15,16], already linking this phenomenon to the presence of defects and SCRs. In pure and doped $BaZrO_3$ (BZO), the formation of SCRs at GBs was studied mainly in the context of proton conduction, both experimentally[17,18] and by means of atomistic calculations[19,20]. In general, SCRs form around GBs as a result of point defect segregation, which originates from differences in the Gibbs free energies of defect formation between the bulk grain interior and GBs[11,14,21,22], where different atomic environments prevail. The knowledge of oxygen vacancy formation energy profiles across samples with GBs is therefore essential to assess the possibility of field-assisted grain growth in a ceramic material. From a computational perspective, detailed information of the functional form of such profiles can be used as input data for a realistic SCR determination based on thermodynamic models. Recently, in a related context, Rauschen and De Souza[23] determined space charge potentials and oxygen vacancy concentrations across a ferroelastic domain boundary in perovskite, $CaTiO_3$, from such oxygen vacancy formation energy profiles.

In our study, we present such formation energy profiles for oxygen vacancies in the three perovskite compounds STO, BTO, and BZO, which are obtained by atomistic calculations. For each of the three compounds, we considered three GB types: the symmetric tilt GBs (STGBs) $\Sigma 5(210)[001]$ and $\Sigma 5(310)[001]$, and the asymmetric tilt GB (ATGB) $(430)[001]||(100)[001]$. Several experimental studies report their occurrence in ceramic samples and describe the stoichiometry and structural details of $\Sigma 5$-type GBs in STO[24]–[29]. In some of these works, supporting atomistic calculations, mainly based on density functional theory, were performed[25,27,29] and reproduced the experimentally observed GB configurations well. The ATGB was investigated in STO, too, both experimentally[30] and theoretically[31].

In our work, we apply atomistic calculations based on rigid-ion potentials to first optimize the GB configurations and then calculate the oxygen vacancy formation energies. For STO, the Buckingham potential with parameters given by Thomas et al. ("Thomas potential")[32] was previously shown to confirm the structure of experimentally observed $\Sigma 3$ GBs[33,34]. It was further used to study oxygen vacancies in low angle $\Sigma 5$ GBs[35]. For BTO, Oyama et al.[36] reported structures derived by a Buckingham potential to be in good agreement with experimental findings, as well. For BZO, a potential of this type was used by Lindman el al.[37] to calculate oxygen vacancy formation energies across symmetric $(11k)[-110]$ GBs with $k=1–8$.

This paper is organized as follows. First, we give a description of the potential model, the cells used for the simulations, the optimization procedure, and the calculation of oxygen vacancy formation energies (Section 2). Next, the optimized GB structures are described and discussed (Section 3.1), for which we then present oxygen vacancy formation energy profiles (Section 3.2). Finally, conclusions are given in Section 4. We provide various computational details about the potentials, the optimization procedure, the supercells, and the fitting method of the energies to obtain the profiles in the Supporting Information (SI).

## 2. Methods and Model

### 2.1. Interatomic Potential

We considered the compounds STO, BTO, and BZO in the cubic perovskite structure, which is an experimentally observed stable crystal phase for each of these three materials. The particles were treated as rigid ions with fixed charges $q_i$, interacting by means of a Coulomb-Buckingham potential:

$$U_{ij}(r) = A_{ij} e^{-r/\rho_{ij}} - \frac{C_{ij}}{r^6} + \frac{q_i q_j}{4\pi\varepsilon_0 r} \tag{1}$$

Here, $r$ denotes the distance between particles $i$ and $j$ and $\varepsilon_0$ is the vacuum permittivity. The potential parameters $A$, $\rho$, and $C$ for ionic pairs, as well as their charges $q$ considered in the respective model were taken from the literature (see SI). For BTO and BZO, those charges are identical to the formal oxidation states of the ions ($Ba^{2+}$, $Ti^{+4}$, $Zr^{4+}$, $O^{2-}$), while for STO, the charges were considered as fitting parameters in the model, leading to $q$(Sr) = +1.84, $q$(Ti) = +2.36, and $q$(Sr) = -1.4[32]. For the three compounds, the potentials reproduce the cubic perovskite ground states with lattice parameters 3.905 Å (STO), 3.992 Å (BTO), and 4.189 Å (BZO). These values are each by about 1% lower compared to the lattice parameters obtained by density functional theory calculations using the generalized gradient approximation[38]. The potentials were also successfully applied to model grain boundaries, where different bonding environments prevail[33]–[37].

### 2.2. Structure models for the optimization of grain boundaries

We constructed supercells with GBs by tilting two bulk regions with respect to each other around the [001] axis by angles corresponding to the orientation relationships, namely by 53.13° for the STGB Σ5(210)[001], by 36.87° for STGB Σ5(310)[001], and, also, by 36.87° in case of the ATGB (430)[001]||(100)[001]. The designation of the tilt axis [001] will be skipped in the following for a better readability. The differently oriented grains were joined at their denoted planes to form the GB (see **Figure 1**). Following the procedure described in Refs. [29,31,36], we identified the optimal configuration of each GB type in each compound by applying relative rigid-body translations (RBTs) in all three directions between the two grains. Structural relaxation at constant volume was performed at each RBT step. The configuration with the lowest GB energy was finally considered as the optimal structure. For the energy calculations, periodic boundary conditions (PBC) were applied in all three directions for all of the systems.

In case of the STBGs Σ5(210) and Σ5(310), the supercells are stoichiometric, and the application of PBC in the direction perpendicular to such a GB requires that the supercells contain two GBs. Since those two GBs are stoichiometrically and structurally identical independent of the RBT state, their GB energy can directly be calculated from the total cell energy. However, in the case of an ATGB, the two GBs in a stoichiometric supercell are in general *not* equivalent to each other. To calculate *unique* GB energies of the ATGB (430)||(100) for different RBT states, we applied the method described by Lee at al.[31], which makes use of slab systems, i.e., of supercells with vacuum at the two ends in the direction perpendicular to the GB, as shown in the bottom panel of Figure 1. For such cells, energy differences *with* and *without* RBT can be calculated. To obtain from those differences total unique GB energies for specific RBT states, reference values of cells without RBT are needed, which must be determined in the same way as for the STGBs, i.e., from supercells without vacuum which contain two equivalent ATGBs. The stoichiometric equivalence of the two ATGBs requires a non-stoichiometric supercell. As mentioned above, the reference supercell must be without RBT, in which case the two ATGBs are also structurally equivalent. While the slab system could in principle be set up stoichiometrically (by omitting the outer lattice planes), we used a non-stoichiometric slab to ensure the same composition of the system of interest and the reference system. We constructed such a slab system, which contains only one GB, with a total vacuum of length $2\Delta_{vac} = 10$ Å. However, this introduces two free surfaces, which could undergo reconstruction during the structural relaxation, and the surface energy would contribute to the total cell energy. Since the surfaces are not of interest in this study, we avoided a potential reconstruction and decreased the influence of the surfaces on the GB structure by fixing a number of layers according to one periodic length in each of the two grains. We provide more details about the cell dimensions and stoichiometries, about the optimization procedure, and about the calculation of the GB energy in the SI.

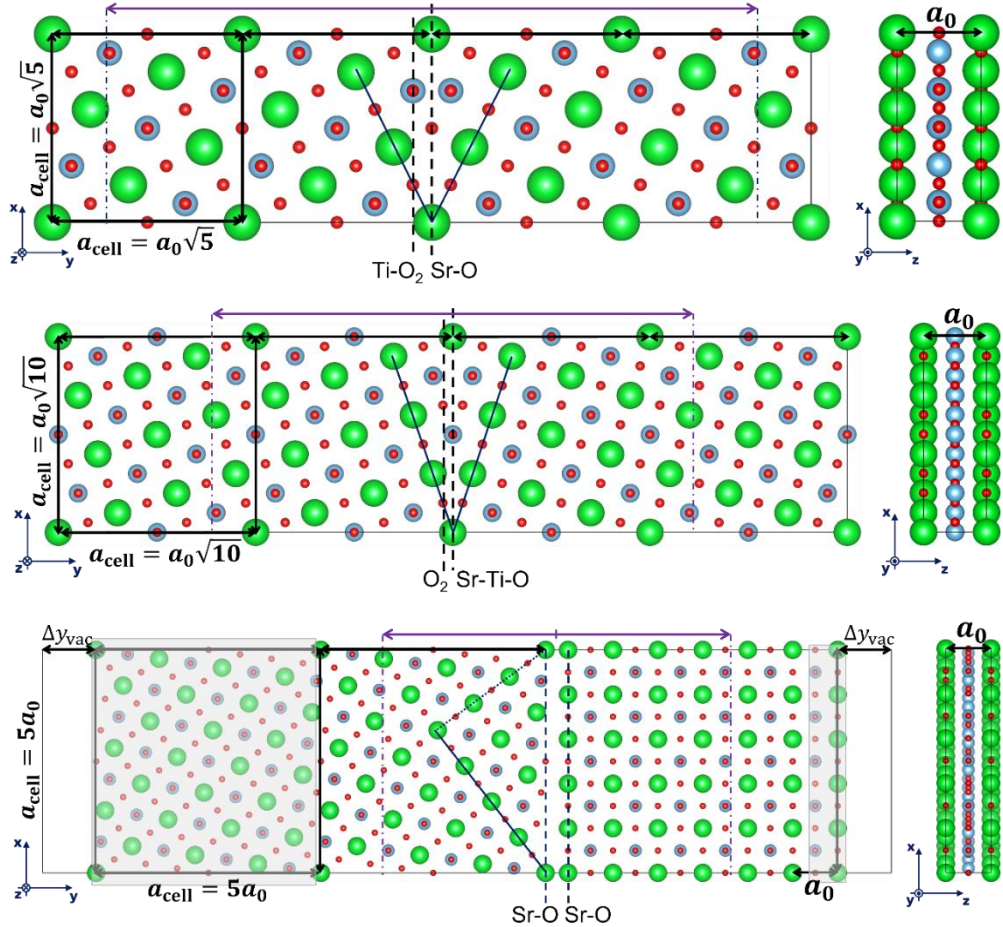

Figure 1. Simulation cells for the optimization of the STGBs $\Sigma5(210)[001]$, (top), $\Sigma5(310)[001]$ (center), and for the ATGB $(430)[001]||(100)[001]$ (bottom), displayed for the example of STO, with Sr atoms colored in green, Ti in blue, and O in red. The stoichiometries of the cells are $Sr_{20}Ti_{20}O_{60}$ (top), $Sr_{40}Ti_{40}O_{120}$ (center), and $Sr_{86}Ti_{80}O_{246}$ (bottom). The corresponding cells for BTO and BZO only differ by their respective lattice constant $a_0$. V-shaped lines indicate the misorientation angles of the grains. The termination planes of the grains are depicted by dashed vertical lines, and their stoichiometries are given. In the ATGB cell, the grains were initially separated by $0.5a_0$ and the atoms in the gray shaded parts of the cell were fixed during the relaxation. The violet double arrows on top of each cell, together with the dash-dotted vertical lines at their ends, indicate regions of length 30 Å, with 15 Å left and right of the centers of the GBs, corresponding to the extensions of the oxygen vacancy formation energy profiles depicted in Figures 4 and 5. The structure figures were prepared with the visualization program VESTA[39].

## 2.3. Oxygen vacancy formation energies

We obtained the formation energies of oxygen vacancies ($V_O$) at all the available oxygen sites within the structurally optimized GB supercells with respect to the formation energy of an oxygen vacancy in the bulk interior ($V_O^{bulk}$) of the respective compound according to:

$$\Delta E^f(V_O) = E(V_O) - E(V_O^{bulk}) + E^{corr} \tag{2}$$

Here, $E$ denotes the total energy of a supercell containing an oxygen vacancy, and $E^{\mathrm{corr}}$ is a correction energy to compensate artifacts of the finite size simulation model. As we described in detail in our previous papers[40,41], lattice planes consisting of charged ions can lead to strong electric fields in the supercells, depending on the stoichiometry of the planes, on the charges of the ions, and on the relative shifts of the lattice planes with respect to each other during the structural relaxation. Such electric fields must be considered as artifacts of the simulation model, since they are not present in realistic systems, where the electric fields can be compensated by stoichiometric changes and an adjustment of the electronic charge density to avoid the "polar catastrophe" of diverging electric fields in macroscopically sized systems[42]–[45]. In case of the supercell of STO containing the ATGB (430)||(100), the non-stoichiometry of the system leads to a net charge in the cell, which is caused by the uncompensated charges on the lattice planes with stoichiometries SrO and $TiO_2$ originating from the ionic charges of the Thomas potential (see Section 2.1). When PBC are applied, such a net charge in the supercell must be compensated, e.g., by a neutralizing background charge density, since it leads to an electric field in the simulation cell. In the rigid ion model, the oxygen ion is negatively charged, therefore the oxygen vacancy acts as a positive point charge located at the position of the extracted ion. The interaction energy of the charged point defect with the artificial field must be corrected, and we followed the procedure described in Ref.[41] to obtain $E^{\mathrm{corr}}$ in Equation (2) for the different GB cells considered in this work. The method is based on analytic functions for the electrostatic potentials and the subtraction of the corresponding energies from the outcome of the energy calculations of the cells containing the charged oxygen vacancies. All the results presented here were calculated with the *General Utility Lattice Program* (GULP)[46].

## 3. Results and Discussion

### 3.1. Grain boundary structures

After setting up the supercells and performing the optimization procedure, we obtained the structures of the STGBs Σ5(210) and Σ5(310) depicted in **Figure 2**. The leftmost images display the initial structures before optimization, which are the same for STO, BTO, and BZO, except for the different lattice constants. For both GBs, the initial structures contain ions of the same types close to each other within the characteristic structural unit (with "kite" shape) formed by the ions at the GBs. The optimization procedure remedies those unfavorable configurations in different ways.

In the final Σ5(210) structure in STO, the two grains left and right of the GB (with respect to the y-direction) are shifted against each other by half a lattice constant in the direction perpendicular to the kite plane (z-direction). This allows the atoms to nearly keep their positions projected onto that plane, except for a slight widening of the kite. There is no such z-shift for the Σ5(210) GB in BTO. Instead, the electrostatic energy of repelling ions is decreased by a larger increase of the kite area and the movement of the central oxygen ion in between the neighboring Ti ions. Note, that in each of the relaxed Σ5(210) structures, the oxygen ions along the red line crossing the kite are alternatingly staggered in z-direction. Compared to BTO, the larger Zr ions in BZO distort the kite structure more strongly. The z-shift between the left and the right grain in STO does not appear in BZO either.

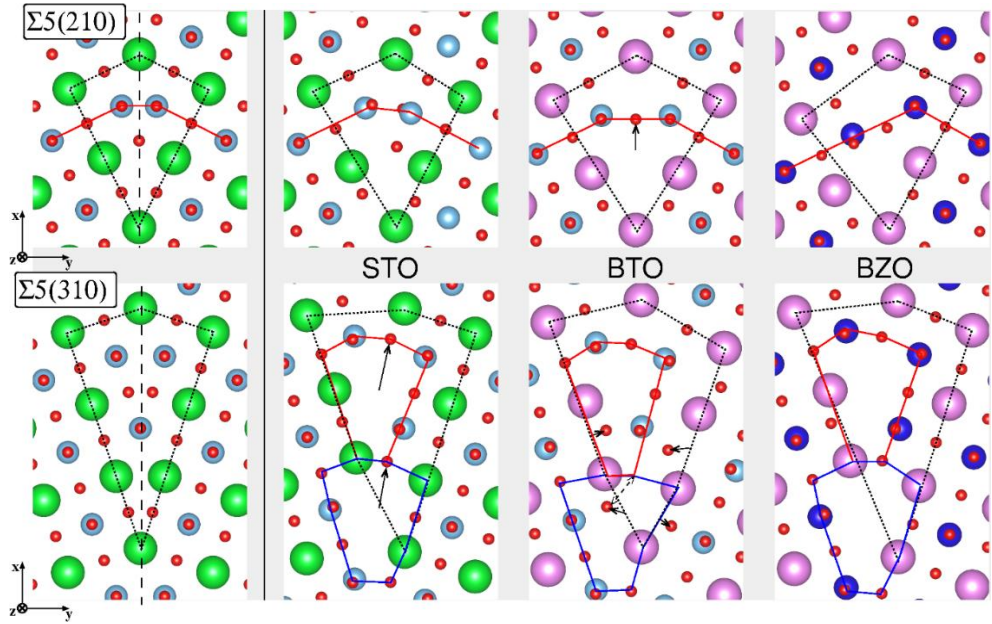

*Figure 2. Two-dimensional representations of the GB structures Σ5(210) (upper pictures) and Σ5(310) (lower pictures) before (leftmost two pictures) and after optimization (pictures to the right of the vertical solid line) for STO, BTO, and BZO. The dashed vertical lines in the two leftmost pictures indicate the mirror planes of the unoptimized GBs. The dotted lines represent the "kite" structural unit characteristic of the GB type. For the Σ5(210) configurations, the red lines indicate a sequence of oxygen ions alternatingly shifted in z-direction by half a lattice constant. For the Σ5(310) configurations, areas enclosed with red and blue lines highlight the "voids" characteristic of this GB type. Exemplary, arrows indicate shifts of atoms with respect to the initial structure (solid) or to final structures of another composition (dashed). Sr atoms are colored in green, Ba in violet, Ti in light blue, Zr in dark blue, and O in red.*

In the initial Σ5(310) configuration, there are more unfavorably arranged neighboring ions of the same type than in the case of Σ5(210). This leads to a stronger deviation of the structures upon relaxation, with the formation of "voids" which are characteristic of this GB type. STO and BZO have a similar final structure. In those systems, one of the two neighboring oxygen ions at the lower cusp of the kite (with respect to the x-direction) moves in between the neighboring Sr/Ba ions, accompanied with an asymmetric distortion of the kite in this region. This creates the voids enclosed by the blue lines in Figure 2. The voids enclosed by the red lines are formed analogously by a movement of an oxygen ion between the neighboring Ti/Zr ions. In BTO, this "upper" void can easily be identified, too, the "lower" void hardly. Instead of the oxygen ion moving between the Ba ions, both Ba-Ba and O-O pairs just increase their respective distance to lower the electrostatic energy. RBTs of the grains against each other in z-direction were not found for the Σ5(310) structures.

For STO, studies were carried out which combined first principles calculations based on density functional theory (DFT) and electron microscopy to analyze the structures of both Σ5(210)[29] and Σ5(310)[27] GBs. The reported stable configurations agree qualitatively with those obtained by the potential used in our work. This applies to BTO, too, as it was already described by Oyama et al.[36], who used the same interaction model. A GB Σ5(210) structure in BZO was optimized by means of DFT by Lindman et al.[47], who report a z-shift as in our final STO configuration. As described above, this state is however not the most favorable one when applying the potential for BZO used in our work. The GB Σ5(310) in BZO was analyzed by DFT, as well, but without a systematic optimization of the structure by applying RBT[48]. Even though they are a strong simplification of the ionic interactions compared to DFT, we consider the empirical Coulomb-Buckingham-type potentials used in this work to reproduce realistic low energy GB structures as acceptably good. Accordingly, we also applied them for optimizing structures of asymmetric GBs, namely the ATGB (430)||(100).

The final ATGB configurations are depicted in **Figure 3**. There are strong similarities between the optimized structures of STO and BTO: in both cases, a relative shift of the two grains in z-direction by half a lattice constant is favored. In the central region of the GB with respect to the x-direction, a Sr/Ba ion moves from its initial position on the terminating (430) plane into a more open region between the grains, which is more pronounced in BTO than in STO. The (100) termination plane is hardly distorted in both compounds. A considerable void remains in the lower x-region. This cannot be filled by the displacement of surrounding ions without substantially increasing the distances between anions and cations and an approaching of alike ions, which in sum would be energetically unfavorable.

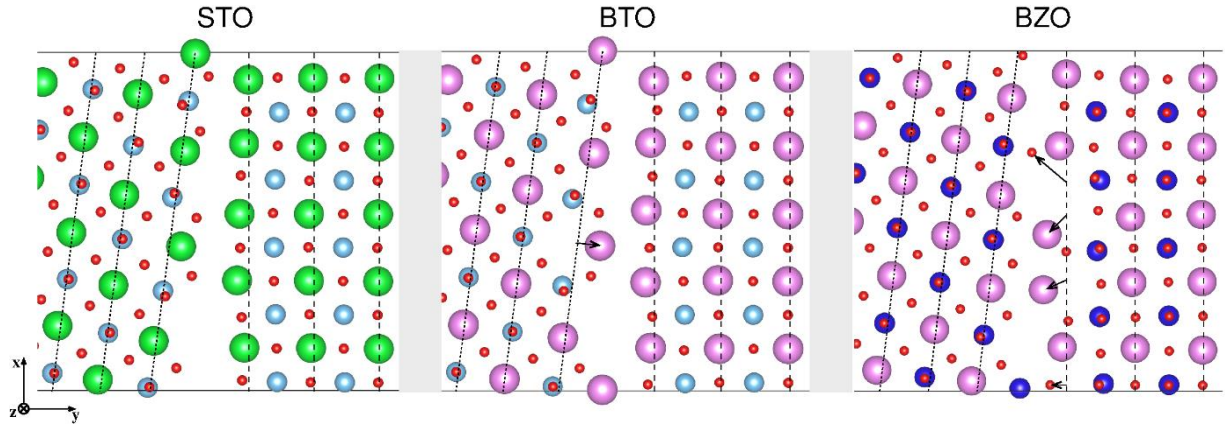

*Figure 3. Two-dimensional representations of the ATGB (430)||(100) structures of STO, BTO, and BZO after optimization. The left grains have the (430) orientation, and the right grains the (100) orientation in each case. Dashed lines indicate the ionic positions on the initially set, not yet relaxed planes. Arrows mark the movements of atoms away from those planes during relaxation. Atoms are colored as in Figure 2.*

The situation in BZO is different, which is primarily due to a different preferred termination of the grain oriented in (430) direction. There is no z-shift between the grains as in STO and BTO, but instead the ions on the termination plane of the (100)-oriented grain are strongly displaced from their initial positions, whereas those on the (430)-oriented grain remain close to their initial positions. By performing DFT calculations to optimize the structures and comparing the results to experiments, Lee et al.[31] showed for the example of STO, that this strong variation in ion displacement on the termination planes depends indeed on the combination of termination planes of the two grains (i.e., SrO-SrO, SrO-TiO$_2$, TiO$_2$-SrO, or TiO$_2$-TiO$_2$), being the least in the SrO-SrO case. For the optimized ATGB structures of STO and BTO displayed in Figure 3, we found this to be the energetically preferred termination combination, and the ionic configurations at the GB are qualitatively similar to the structure shown in Ref. [31] for this case.

### 3.2. Oxygen vacancy formation energies

For each of the three GB types in the three perovskite compounds, we calculated the formation energies of oxygen vacancies at all the oxygen sites in the supercell according to Equation (2). To make sure that finite size effects do not influence the results, the supercells were increased in z-direction from $1a_0$ (which was used for the GB optimization, see Figure 1) to $3a_0$. In addition, in case of the GBs Σ5(210) and Σ5(310), the length of the bulk in y-direction was increased by a factor of 2 to decrease the influence of the GBs on the vacancy formation energies in the bulk region. In case of the ATGB, the y-dimension was kept at the values used for optimization as they turned out to be already large enough to exclude such a finite-size effect.

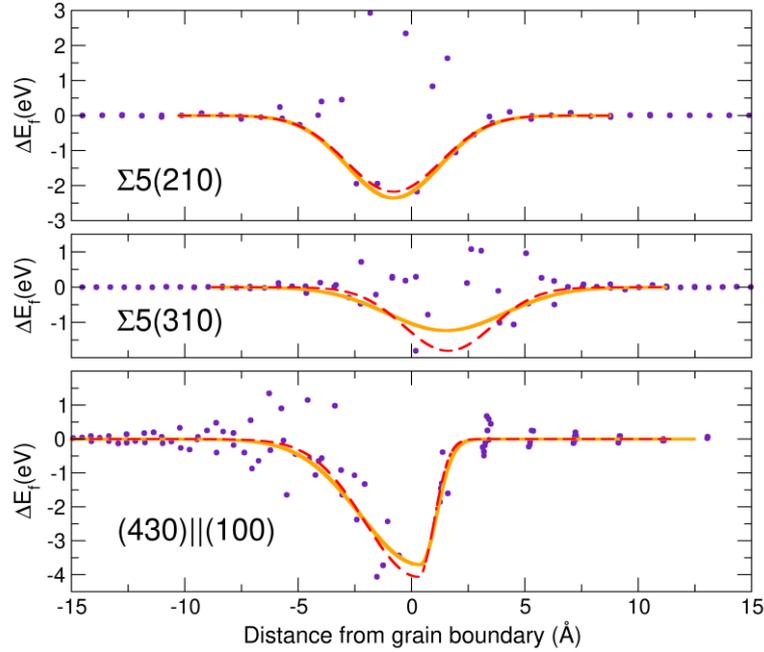

*Figure 4. Relative oxygen vacancy formation energies with respect to the bulk crystal interior within regions of ±15 Å around the cores of the different GB types (indigo-colored points) in STO. Orange solid lines are the averaged profiles from Gaussian fits of the negative values in the considered regions following "method (1)" (see description in the text), where the depth is treated as a fit parameter, and red dashed lines were obtained by fitting with "method (2)", where the depth is set to the lowest data point.*

For STO, the results for the 3 GB types in STO are shown in **Figure 4** as a function of the distance from the GB in y-direction, i.e., perpendicular to the GB plane, in a region of ±15 Å around the GB. The formation energy values in the bulk parts of the supercells approach their bulk value within 5–10 Å. This indicates that the simulation cells were chosen large enough to exclude an influence of the second GB present in the cell in case of the STGBs, and of the free surfaces in the case of the ATGB.

In the core regions of the GBs, the formation energy values strongly deviate from their respective bulk value, both in positive and negative directions. Considering the Arrhenius dependence between defect formation energies and concentrations valid in the dilute limit, oxygen vacancies will form with lower probability than in the bulk if the formation energies are positive. At such sites, the oxygen ions are in an even more favorable bonding environment than in the perfect structure, which stabilizes the GB. On the other hand, there are sites with negative formation energies with respect to the bulk grain interior, which, depending on the ambient temperature during the phase formation in the synthesis process, would be vacant with concentrations that are orders of magnitude above the bulk concentration. We obtained averaged profiles of negative vacancy formation energies across the GBs by applying Gaussian fits. This is justified, since on average, the unfavorable oxygen sites are located deeper in the GB core regions, which is most pronounced in case of the asymmetric GB. In the latter case, we performed "asymmetric Gaussian fits", i.e., we fitted two different, but connected Gaussians for the regions left and right of the GB, to more accurately account for the strongly differing behavior of the formation energy values caused by the much more different structural environment as in the case of the symmetric GBs.

For all considered GBs, we performed the fits of the negative formation energies in two different ways: (1) by treating the depth of the Gaussian as a fit parameter, and (2) by fixing the depth at the value of the lowest formation energy value. The fit lines obtained by method (1) give us more insight into the average behavior of the formation energies across the GBs, which we can compare for different GB types and stoichiometries. Individual outliers in the data points are not reflected by those lines. Method (2) on the other hand enforces the lowest values to be covered by the fit lines. Profiles of this type are more relevant from the practical perspective as input for the determination of space charge regions by thermodynamic models[20]. As shown in Figure 4, the difference in the fit lines obtained by methods (1) and (2) are rather insignificant for the GBs Σ5(210) and for the ATGB, while one comparatively low-lying data point within the GB Σ5(310) drags the corresponding fit line from method (2) further down. The fit parameters and the standard deviations of the fits are summarized in the SI for both methods.

In STO, the formation energies are the least spread around zero in the GB Σ5(310) configuration. This is related to the very regular structure of the GB with the two characteristic voids in its core as shown in Figure 2. The spread of formation energies is strongest in case of the ATGB, with many rather unfavorable sites for oxygen atoms, mainly located in the termination region of in the grain oriented in (430) direction. In STO containing symmetric GBs of the Σ3 type, formation energies of oxygen vacancies were calculated applying DFT[49] and the potential used in this work[34]. In both cases, negative values with respect to the bulk grain interior were found in the GB cores, which is consistent to an experimental study by Zhang et al.[50], who reported the existence of oxygen vacancies in the Σ3 GBs.

To estimate the concentrations of oxygen vacancies in the GB regions, one can use the relative values of formation energies at the GBs with respect to the bulk, as derived in this study, in combination with the Arrhenius law, if information about the defect chemistry in the bulk is available, either from experiments[51]–[53] or from *ab initio* calculations[54,55]. In the latter case, one typically determines the absolute value of the formation energy of a bulk reference site, which however is not a fixed number but in general depends on the oxygen partial pressure and temperature. Since we are dealing with charged oxygen vacancies, the bulk value of the formation energy additionally depends on the Fermi level. For oxygen rich conditions (high oxygen partial pressure), Ricca et al.[55] obtained from DFT calculations formation energies of bulk oxygen vacancies in STO from about 0.5 eV for Fermi levels at the valence band maximum to about 6 eV for Fermi levels at the conduction band minimum. Considering the values given in Figure 4, low Fermi levels would correspond to negative absolute formation energies in the GBs leading to very high oxygen vacancy concentrations and unstable GB structures. However, high concentrations of this donor-like defect would in turn rise the Fermi level. Therefore, values above the middle of the band gap can be assumed, for which, according to Ref. [55], the oxygen vacancy formation energies would have positive absolute values even in the core of the ATGB, and the structure would remain stable.

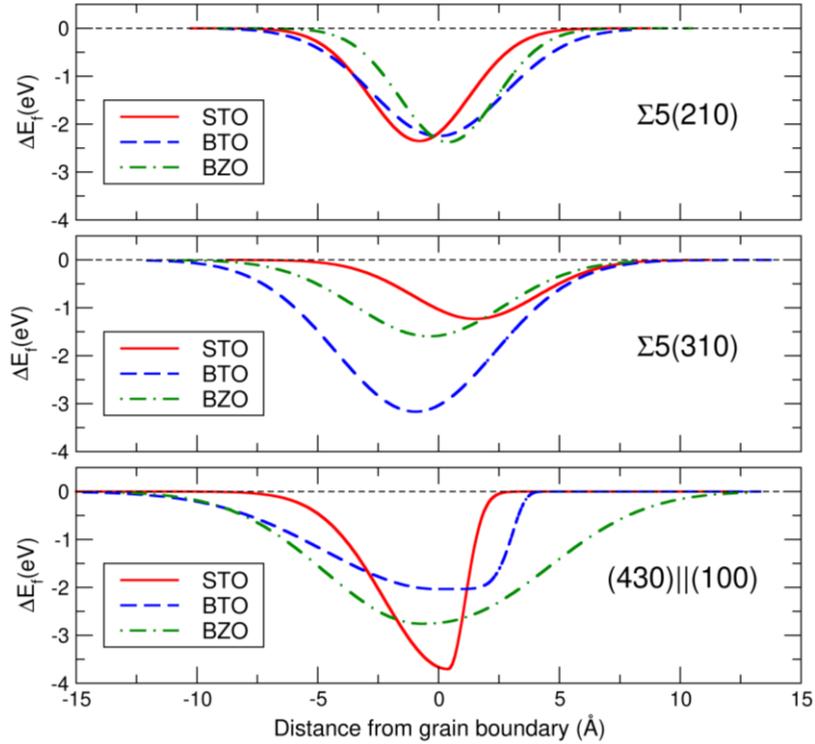

*Figure 5. Gaussian profiles [resulting from method (1)] of the negative oxygen vacancy formation energies with respect to values in the bulk grain interior for the three GB types in the three considered perovskite compounds.*

**Figure 5** displays the Gauss-fitted profiles [from method (1), see description above] of negative relative oxygen vacancy formation energies across the GBs for STO, BTO, and BZO. The underlying energy values for the latter two can be found in the SI, together with the fit lines obtained from methods (1) and (2), and with the corresponding fit parameters. As apparent from Figure 5, in the case of the ATGB, STO and BTO have rather different averaged profiles in terms of width and depth, even though their GB structures appear to be quite similar (see Figure 3). The strong quantitative differences must therefore originate from the different chemistries of the two materials, which are only implicitly accounted for by the different parameterizations of the empirical Coulomb-Buckingham potentials. However, there are qualitative similarities, namely the smooth decrease in the (430)-oriented grain when approaching the GB core and, following this direction, the steeper ascent when entering the (100)-oriented region. Compared to STO and BZO, the profile of BTO across the ATGB is also qualitatively different, which relates to the strongly different structure displayed in Figure 3 and discussed in Section 3.1.

For the GB $\Sigma5(310)$, the chemical influence on the profiles seems to cancel out between STO and BZO, and the curves are qualitatively and quantitatively similar, which is consistent to the similarity of their GB structures displayed in Figure 2. The comparatively more displaced oxygen sites in the corresponding BTO configuration are less favorable and have therefore lower vacancy formation energies, which is apparent in the profile. A clear distinction between the influences of chemistry and structure is not possible for the GB $\Sigma5(210)$, where the profiles have similar shapes, but the GB configurations have different characteristics.

If needed as useful input for thermodynamic calculations of space charge regions, the fit lines corresponding to those shown in Figure 5 but obtained from the alternative Gauss fits [method (2)] can be found in the SI.

## 4. Conclusions

In the context of abnormal grain growth in perovskite ceramics in electric field assisted sintering, we investigated by means of atomistic simulations how the formation energy of oxygen vacancies is related to grain-boundary structures in three prominent perovskite compounds, STO, BTO and BZO. We presented optimized grain boundary structures of STGBs $\Sigma 5(210)$, $\Sigma 5(310)$ and of the ATGB $(430)||(100)$, and profiles of oxygen vacancy formation energies across those GBs.

Even though the optimization procedure was conducted in the same way and starting from the same initial structures, the final, lowest energy GB configurations of the same type exhibit a variety of structural features for the different chemical compositions. Within the configurational constraints, the atoms find different ways to arrange such that the energy is minimized.

Structural variabilities of the grain boundaries can qualitatively be spotted in the fitted Gaussian profiles of the calculated formation energies of oxygen vacancies, with a more [(430)||(100) in STO and BTO] or less [$\Sigma 5(310)$ in STO and BZO] pronounced influence of the chemistry. Considering the obtained average profiles, structural and chemical effects may also compensate each other, as visible in the qualitatively and quantitatively very similar profiles across $\Sigma 5(210)$ in the three compounds.

Altogether, all three GB types in all three considered perovskite materials exhibit negative formation energies of oxygen vacancies, with the corresponding sites acting as trap sites for this point-defect species. The lower the energy, the more likely the oxygen sites in the GB core are vacant, which gives rise to larger extended space charge regions. In the case of STO, this is consistent with the assumption underlying the thermodynamic model proposed by Rheinheimer et al.[10], which explains their experimental findings of electric field-assisted grain growth for this material. BTO and BZO, for which there are no reports of corresponding experiments so far, can therefore be considered as potential materials for such a grain-growth behavior, too.

Due to a lack of detailed microscale information on the formation and distribution of point defects at interfaces, often just a single value for the relative oxygen vacancy formation energy in the GB is taken into account in the thermodynamic models. Using the complete profiles presented in this work, in combination with analog curves for cation vacancies, will therefore help to come to a more realistic description of space charge regions in thermodynamic models for studying grain growth phenomena in ceramic microstructures.

**Supporting Information** is provided in a separate document

**Acknowledgements**

This work was funded by the German Research Foundation (DFG); Grants No. MR22/6-1 and EL155/31-1 within the priority program 'Fields Matter' (SPP 1959). Computations were carried out on the bwUniCluster computer system of the SteinbuchCentre of Computing (SCC) of the Karlsruhe Institute of Technology (KIT), funded by the Ministry of Science, Research, and Arts Baden-Württemberg, Germany, and by the DFG.

# Supporting Information

for

# Formation energy profiles of oxygen vacancies at grain boundaries in perovskite-type electroceramics

Daniel Mutter[1], Cong Tao[1*], Daniel F. Urban[1,2], Christian Elsässer[1,2]

[1]Fraunhofer IWM, Wöhlerstraße 11, 79108 Freiburg, Germany
[2]Freiburg Materials Research Center (FMF), University of Freiburg, Stefan-Meier-Straße 21, 79104 Freiburg, Germany

*Corresponding author
 E-mail: cong.tao@iwm.fraunhofer.de

This document contains supporting information about the potential parameters of the Coulomb-Buckingham potential (Section SI-1), about the grain boundary optimization, rigid body translations, and the calculation of the grain boundary energy (Section SI-2), and about the oxygen vacancy formation energy values, and the fitting procedure and parameters (Section SI-3).

**SI-1. Potential parameters of the Coulomb-Buckingham potential**

**Table SI-1** lists the parameters of the Coulomb-Buckingham pair potential applied in this work [Equation (SI-1)] to model SrTiO$_3$ (STO), BaTiO$_3$ (BTO), and BaZrO$_3$ (BZO):

$$U_{ij}(r) = A_{ij}e^{-r/\rho_{ij}} - \frac{C_{ij}}{r^6} + \frac{q_i q_j}{4\pi\varepsilon_0 r} \tag{SI-1}$$

*Table SI-1. Parameters of the Coulomb-Buckingham potential*

| Compound | Ion pair | $A$ (eV) | $\rho$ (Å) | $C$ (eV Å$^6$) | $r_{\text{cut-off}}$ (Å) |
|---|---|---|---|---|---|
| STO[1] | Sr$^{1.84+}$ – O$^{1.4-}$ | 1769.51 | 0.319894 | 0.000 | 20 |
| | Ti$^{2.36+}$ – O$^{1.4-}$ | 14567.40 | 0.197584 | 0.000 | |
| | O$^{1.4-}$ – O$^{1.4-}$ | 6249.17 | 0.231472 | 0.000 | |
| BTO[2]–[4] | Ba$^{2+}$ – O$^{2-}$ | 905.700 | 0.39760 | 0.000 | 10 |
| | Ti$^{4+}$ – O$^{2-}$ | 2179.122 | 0.30384 | 8.986 | |
| | O$^{2-}$ – O$^{2-}$ | 9547.960 | 0.21916 | 32.000 | |
| BZO[5] | Ba$^{2+}$ – O$^{2-}$ | 931.700 | 0.3949 | 0.000 | 6 |
| | Zr$^{4+}$ – O$^{2-}$ | 985.869 | 0.3760 | 0.000 | |
| | O$^{2-}$ – O$^{2-}$ | 22764.300 | 0.1490 | 27.890 | |

Note that for STO, the charges are not the formal charges corresponding to the chemical oxidation states, and the parameters $C_{ij}$ are set to zero ("Thomas potential").

**SI-2. Grain boundary optimization, rigid body translations, and the grain boundary energy**

The procedure to optimize the GBs follows Refs. [6]–[8].

Rigid body translations were performed by shifting the two grains in x- and z-direction with respect to each other ($\delta_x$, $\delta_z$), and by separating them in y-direction ($\delta_y$). At each step, structural relaxation was performed at constant volume, and the total energy was calculated. We applied periodic boundary conditions (PBC) in all three directions for the optimization of the STGB structures. This leads to two equivalent GBs in the supercell. The GB energy $\gamma$ for an RBT state $\vec{t} = (\delta_x, \delta_y, \delta_z)$ is thus given as:

$$\gamma(\vec{t}) = \frac{E_{\text{GB}}^{\text{PBC}}(\vec{t}) - N_{\text{f.u.}} E_{\text{f.u.}}^{\text{bulk}}}{2 A_{\text{GB}}} \tag{SI-2}$$

$E_{\text{GB}}^{\text{PBC}}(\vec{t})$ denotes the total energy of the system with GB using PBC. $E_{\text{f.u.}}^{\text{bulk}}$ is the total energy of one formula unit (f.u.) of the compound in the perfect bulk phase, $N_{\text{f.u.}}$ is the number of f.u.'s in the supercell with GB, and $A_{\text{GB}}$ is the area of the GB [$A_{\text{GB}} = a_0^2\sqrt{5}$ for Σ5(210), $A_{\text{GB}} = a_0^2\sqrt{10}$ for Σ5(310)].

For an ATGB scenario, the two GBs which exist in the cell when PBC are applied in the direction normal to the GB (here, the y-direction) are in general not equivalent to each other except for special RBT states. In our construction, this is the case for $\vec{t} = \vec{0}$, i.e., for the cell without RBT between the grains. For $\vec{t} \neq \vec{0}$, a GB energy referring to a unique GB configuration can therefore not be obtained from Equation (SI-2). To optimize the ATGB with respect to the GB energy, we constructed a slab system with vacuum in y-direction, which contains only one GB. For this slab cell with total energy $E_{\text{GB}}^{\text{slab}}$, we obtained the quantity $\Delta\gamma(\vec{t})$ as the difference of the GB energies with and without the application of RBT:

$$\Delta\gamma(\vec{t}) = \frac{E_{\text{GB}}^{\text{slab}}(\vec{t}) - E_{\text{GB}}^{\text{slab}}(\vec{t} = \vec{0})}{A_{\text{GB}}} \tag{SI-3}$$

By this procedure, the surface energies cancel out. Here, $A_{\text{GB}} = 5a_0^2$. To obtain the total GB energy, its value for $\vec{t} = \vec{0}$ must be added, $\gamma(\vec{t}) = \gamma(\vec{t} = \vec{0}) + \Delta\gamma(\vec{t})$, which can be calculated from the cell without vacuum and without fixed layers by using PBC, according to:

$$\gamma(\vec{t} = \vec{0}) = \frac{E_{\text{GB}}^{\text{PBC}}(\vec{t} = \vec{0}) - N_{\text{f.u.}} E_{\text{f.u.}}^{\text{bulk}} \pm \Delta N_c \mu_c}{2 A_{\text{GB}}} \tag{SI-4}$$

This expression is equivalent to Equation (SI-2) except for the term $\pm \Delta N_c \mu_c$. The stoichiometry of a cell containing an ATGB may differ from an integer number of formula units (f.u.) of the considered material, which is not the case for STGB cells (see detailed explanation in the next paragraph below). Such a discrepancy must be taken into account in the energy balance by adding or subtracting $\Delta N_c$ formula units of a specific compound (c) with energy per f.u. described by the chemical potential $\mu_c$. The compound c and the number $\Delta N_c$ depend on the terminations of the grains. For example, an SrO-SrO termination in a cell of STO with ATGB (430)||(100) can lead to a cell stoichiometry of $Sr_{86}Ti_{80}O_{246}$, corresponding to $\Delta N_c = 6$ for c = SrO.

The origin of the non-stoichiometry can be explained as follows: two stoichiometrically identical ATGBs of the type (430)||(100) in a supercell, which is required for the application of Equation (SI-4), can only be realized if the cell is non-stoichiometric. This is because both of the alternating individual lattice planes in both, the (430)- and the (100)-oriented grains, are non-stoichiometric. For the example of STO, the planes in the (430)-oriented grain have the stoichiometries SrO ("$A$") and TiO$_2$ ("$a$"), and the planes in the (100)-oriented grain have the stoichiometries Sr$_5$O$_5$ ("$B$") and Ti$_5$O$_{10}$ ("$b$"). A cell with two stoichiometrically identical grain boundaries in the case of PBC in the direction perpendicular to the grain boundary requires a sequence of planes such as ...$B|Aa...AaABb...BbB|A$..., where the grain boundary termination pair $AB$ (SrO–Sr$_5$O$_5$) was exemplarily chosen, and $|...|$ delimits the supercell. In total, the stoichiometry of this cell is (SrTiO$_3$)$_n$ + Sr$_6$O$_6$ (with $n \geq 6$), i.e., it is non-stoichiometric. Omitting the outer layers SrO and Sr$_5$O$_5$ would lead to a stoichiometric cell, but to the two different grain boundaries (SrO–Sr$_5$O$_5$) and (TiO$_2$–Ti$_5$O$_{10}$).

The chemical potential $\mu_c$ appearing in Equation (SI-4) can generally be expressed as:

$$\mu_c = \mu_c^{(0)} + \Delta\mu_c \tag{SI-5}$$

Here, $\mu_c^{(0)}$ denotes the total energy per formula unit of the ground state structure of the compound c. For the study of different terminations of ATGB structures in three considered perovskite compounds, we needed to determine the values of $\mu_c^{(0)}$ for SrO, BaO, TiO$_2$, and ZrO$_2$ using the respective potentials (see Section SI-1). The values of $\Delta\mu_c$ are not fixed but can vary in a range corresponding to *rich* or *poor* conditions of the precursor compound c in a hypothetical solid-state synthesis of the material[9]. For example, for STO, the reaction would be SrO + TiO$_2$ → SrTiO$_3$ with a formation energy

$$\Delta E_{STO}^{form} = \Delta\mu_{SrO} + \Delta\mu_{TiO_2} \tag{SI-6}$$

Having $\Delta E_{STO}^{form}$ obtained from the pair potential, the value of $\Delta\mu_{SrO}$ ranges between 0 eV (SrO-rich conditions) and $\Delta E_{STO}^{form}$ (SrO-poor / TiO$_2$-rich conditions).

For all three GB types in the three considered compounds, the optimization of the GB configuration with respect to the GB energy was performed by applying RBT states $\vec{t}$ on a grid with step sizes of 0.1$a_{cell}$ in x-direction for the STGB cells and 0.02$a_{cell}$ = 0.1$a_0$ for the ATGB cells, 0.1 Å in y-direction, and 0.5$a_0$ in z-direction. At each grid point, structural relaxation of the cell was performed at constant volume, and the GB energy was calculated. The final RBT states after the relaxation were obtained by comparing the shifts of two sites deep inside the bulk parts of the two grains with each other. The values are given in Table SI-1.

For the ATGB configurations, the 4 possible combinations of termination planes were tested, namely, SrO-SrO, SrO-TiO$_2$, TiO$_2$-SrO, and TiO$_2$-TiO$_2$ (and analogous combinations for BTO and BZO) with the first composition referring to the termination of the grain oriented in (430)-direction and the second composition referring to the termination of the grain oriented in (100)-direction. Considering in addition the ranges of chemical potentials of the corresponding compounds according to Equations (SI-5) and (SI-6), the terminations leading to the overall lowest GB energies were SrO-SrO for STO, BaO-BaO for BTO, and ZrO$_2$-BaO for BZO. The corresponding ranges of formation energies between Sr(Ba)O-rich and Sr(Ba)O-poor conditions are given in **Table SI-2**.

*Table SI-2. Final RBT states and corresponding GB energies*

| Compound | GB type | $\delta_x$ ($a_{cell}$) | $\delta_y$ (Å) | $\delta_z$ ($a_0$) | $\gamma$ (J/m²) |
|---|---|---|---|---|---|
| STO | Σ5(210) | 0.00 | 0.35 | 0.50 | 1.54 |
|  | Σ5(310) | 0.33 | 1.14 | 0.00 | 1.81 |
|  | (430)\|\|(100) | 0.05 | 0.77 | 0.50 | 1.16 – 1.46 |
| BTO | Σ5(210) | 0.00 | 1.36 | 0.00 | 1.03 |
|  | Σ5(310) | 0.09 | 1.68 | 0.00 | 1.56 |
|  | (430)\|\|(100) | 0.08 | 1.17 | 0.50 | 1.02 – 1.09 |
| BZO | Σ5(210) | −0.06 | 1.12 | 0.00 | 1.37 |
|  | Σ5(310) | 0.33 | 1.16 | 0.00 | 1.52 |
|  | (430)\|\|(100) | 0.08 | 0.68 | 0.00 | 1.89 – 2.20 |

**Table SI-3** lists the dimensions and stoichiometries of the supercells with optimized GBs Σ5(210), Σ5(310), and (430)||(100) in STO, BTO, and BZO.

*Table SI-3. Cell dimensions and stoichiometries*

| Compound | GB-type | Cell parameters (Å) | | | Number of ions | | |
|---|---|---|---|---|---|---|---|
|  |  | $a_{cell}$ | $b_{cell}$ | $c_{cell}$ | Sr, Ba | Ti, Zr | O |
| STO | Σ5(210) | 8.732 | 35.975 | 3.905 | 20 | 20 | 60 |
|  | Σ5(310) | 12.348 | 51.555 | 3.905 | 40 | 40 | 120 |
|  | (430)\|\|(100) | 19.525 | 75.233 | 3.905 | 86 | 80 | 246 |
| BTO | Σ5(210) | 8.922 | 38.408 | 3.992 | 20 | 20 | 60 |
|  | Σ5(310) | 12.716 | 53.830 | 3.992 | 40 | 40 | 120 |
|  | (430)\|\|(100) | 19.960 | 77.069 | 3.992 | 86 | 80 | 246 |
| BZO | Σ5(210) | 9.367 | 39.708 | 4.189 | 20 | 20 | 60 |
|  | Σ5(310) | 13.247 | 55.307 | 4.189 | 40 | 40 | 120 |
|  | (430)\|\|(100) | 20.945 | 79.819 | 4.189 | 85 | 81 | 247 |

**SI-3. Oxygen vacancy formation energy profiles and fitting**

For the three GBs considered in this work, **Figure SI-1** depicts the calculated oxygen vacancy formation energies in BTO, and **Figure SI-2** those in BZO, in both cases relative to a respective bulk value, together with Gaussian fits of the negative values.

For fitting the oxygen vacancy formation energies in the cells containing the STGBs Σ5(210) and Σ5(310), we applied Gauss functions of the form

$$f(y) = Ae^{-\frac{(y-B)^2}{2C^2}}.$$

Two different methods were applied: In method (1), $A$, $B$, and $C$ were treated as fit parameters. In method (2), only $B$ and $C$ were treated as fit parameters, and $A$ was set to the lowest formation energy value. Fit lines resulting from method (1) provide a better average across all of the points, while those resulting from method (2) are more useful as input data for thermodynamic space charge calculations, where the lowest values need to be covered by the profile.

In case of the ATGB (430)||(100), an asymmetric Gauss-like function was used:

$$f(y) = \begin{cases} Ae^{-(\frac{y-B}{C})^D} & : y \leq B \\ Ae^{-(\frac{y-B}{C'})^{D'}} & : y > B \end{cases}$$

with different fit parameters $C$, $C'$ for the two regions separated at $y = B$. $B$ was not treated as a fit parameter in this case but was chosen approximately in the center of the GB. The exponents $D$ and $D'$ were additionally introduced as fit parameters. Also in this case, methods (1) and (2) were applied for treating the parameter $A$ as described above.

The values of the fit parameters are listed in **Table SI-4** [method (1)] and **Table SI-5** [method (2)], together with the standard errors of the fits.

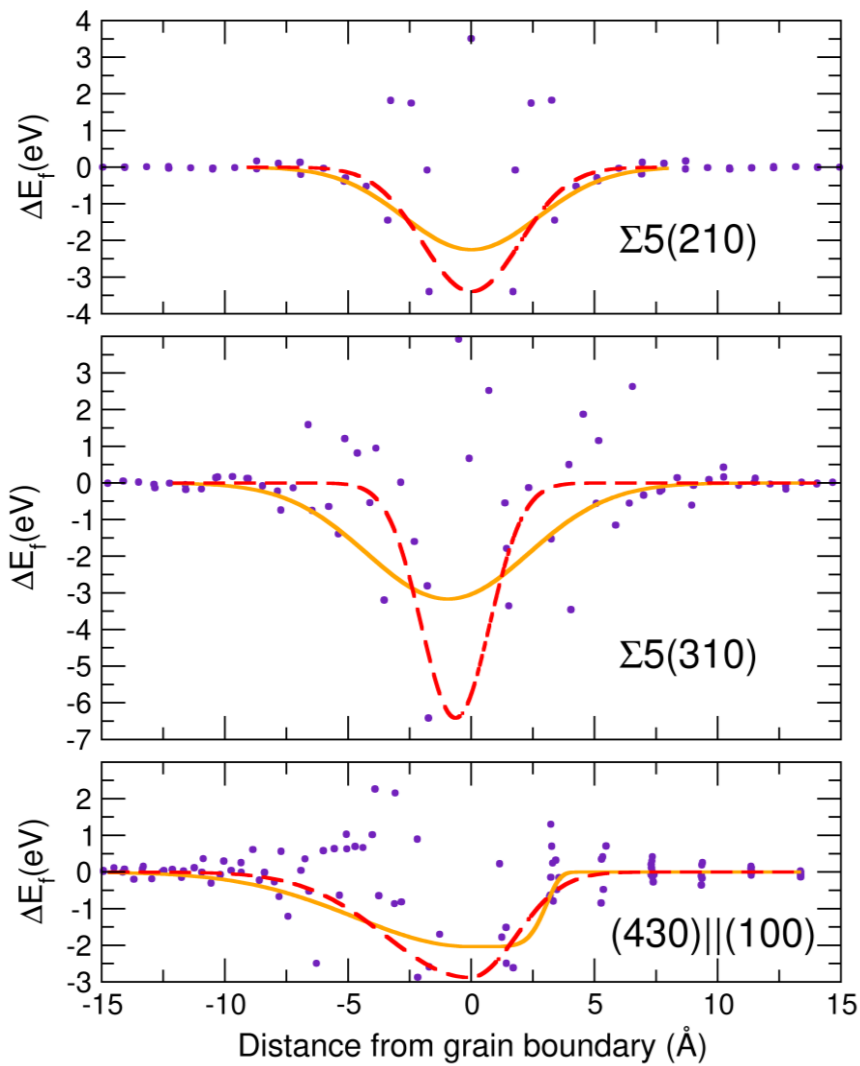

*Figure SI-6. Relative oxygen vacancy formation energies with respect to the bulk value within regions of $\pm 15$ Å around the cores of the different GB types (indigo-colored points) in **BTO**. Orange lines are the averaged profiles from Gaussian fits of the negative values in the GB cores using method (1), where the depth is treated as a fit parameter, and red dashed lines were obtained by fitting with method (2), where the depth is determined by the lowest data point.*

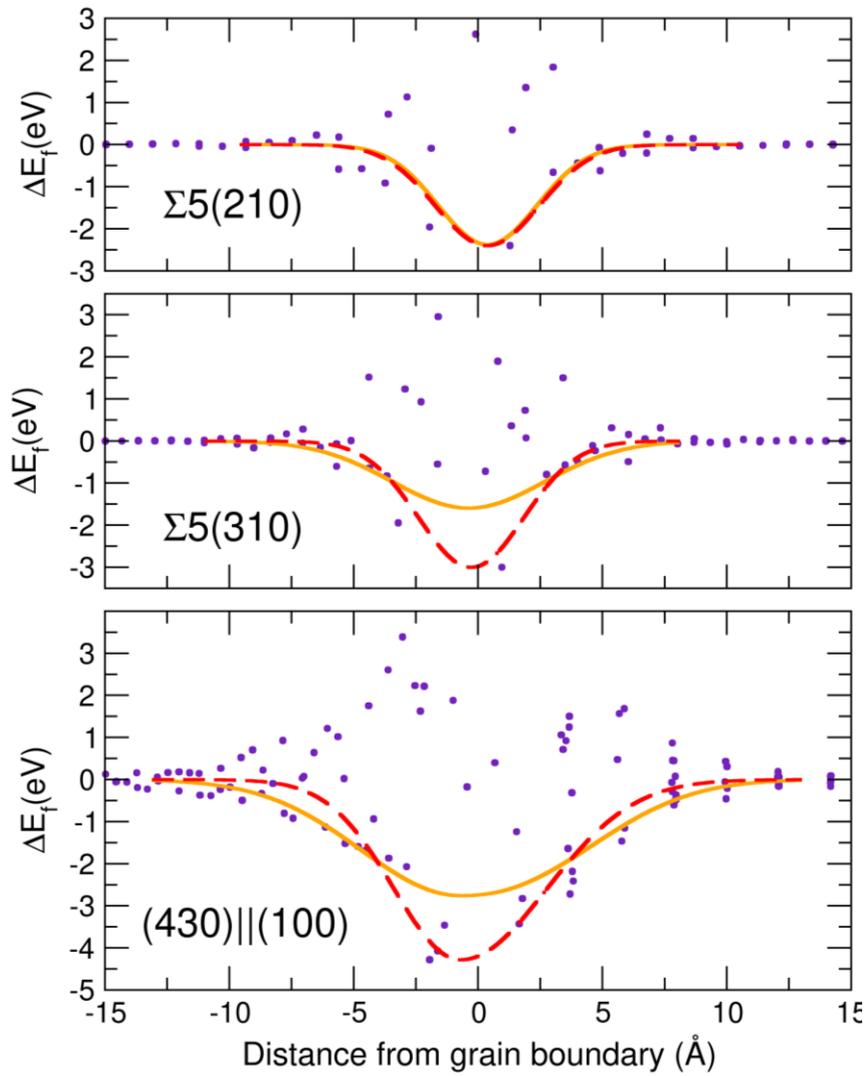

*Figure SI-7. Relative oxygen vacancy formation energies with respect to the bulk value within regions of ±15 Å around the cores of the different GB types (indigo-colored points) in **BZO**. Orange lines are the averaged profiles from Gaussian fits of the negative values in the GB cores using method (1), where the depth is treated as a fit parameter, and red dashed lines were obtained by fitting with method (2), where the depth is determined by the lowest data point.*

Table SI-4. *Values of the fit parameters of the negative relative oxygen vacancy formation energies as a result of **method (1)**, and the standard error of the regressions ($\sigma$).*

| Compound | GB-type | $A$ (eV) | $B$ (Å) | $C$ (Å) | $C'$ (Å) | $D$ | $D'$ | $\sigma$ (eV) |
|---|---|---|---|---|---|---|---|---|
| STO | Σ5(210) | −2.35 | 35.46 | 2.05 | — | — | — | 0.12 |
| STO | Σ5(310) | −1.23 | 50.31 | 2.57 | — | — | — | 0.37 |
| STO | (430)\|\|(100) | −3.58 | 45.81 | 3.68 | 1.01 | 2.00 | 2.00 | 0.43 |
| BTO | Σ5(210) | −2.25 | 37.05 | 2.72 | — | — | — | 0.81 |
| BTO | Σ5(310) | −3.16 | 52.10 | 3.29 | — | — | — | 0.98 |
| BTO | (430)\|\|(100) | −2.03 | 46.63 | 6.59 | 3.19 | 2.00 | 2.52 | 0.46 |
| BZO | Σ5(210) | −2.38 | 39.91 | 2.00 | — | — | — | 0.42 |
| BZO | Σ5(310) | −1.59 | 54.57 | 3.06 | — | — | — | 0.51 |
| BZO | (430)\|\|(100) | −2.76 | 47.33 | 5.64 | 6.78 | 2.00 | 2.31 | 0.67 |

Table SI-5. *Values of the fit parameters of the negative relative oxygen vacancy formation energies as a result of **method (2)**, and the standard error of the regressions ($\sigma$).*

| Compound | GB-type | $A$ (eV) | $B$ (Å) | $C$ (Å) | $C'$ (Å) | $D$ | $D'$ | $\sigma$ (eV) |
|---|---|---|---|---|---|---|---|---|
| STO | Σ5(210) | −2.17 | 35.47 | 2.08 | — | — | — | 0.13 |
| STO | Σ5(310) | −1.81 | 50.41 | 1.96 | — | — | — | 0.42 |
| STO | (430)\|\|(100) | −4.06 | 45.81 | 3.31 | 0.95 | 2.00 | 2.00 | 0.44 |
| BTO | Σ5(210) | −3.39 | 37.05 | 1.98 | — | — | — | 0.87 |
| BTO | Σ5(310) | −6.42 | 52.42 | 1.38 | — | — | — | 1.10 |
| BTO | (430)\|\|(100) | −2.88 | 46.63 | 4.81 | 2.66 | 2.00 | 2.00 | 0.52 |
| BZO | Σ5(210) | −2.40 | 39.90 | 2.08 | — | — | — | 0.43 |
| BZO | Σ5(310) | −3.00 | 54.67 | 2.12 | — | — | — | 0.70 |
| BZO | (430)\|\|(100) | −4.28 | 47.33 | 3.76 | 4.93 | 2.00 | 2.00 | 0.85 |

Figure SI-3 shows a comparison of the fit lines obtained for STO, BTO, and BZO for the two STBGs and the ATGB considered in this work, as resulting from method (2).

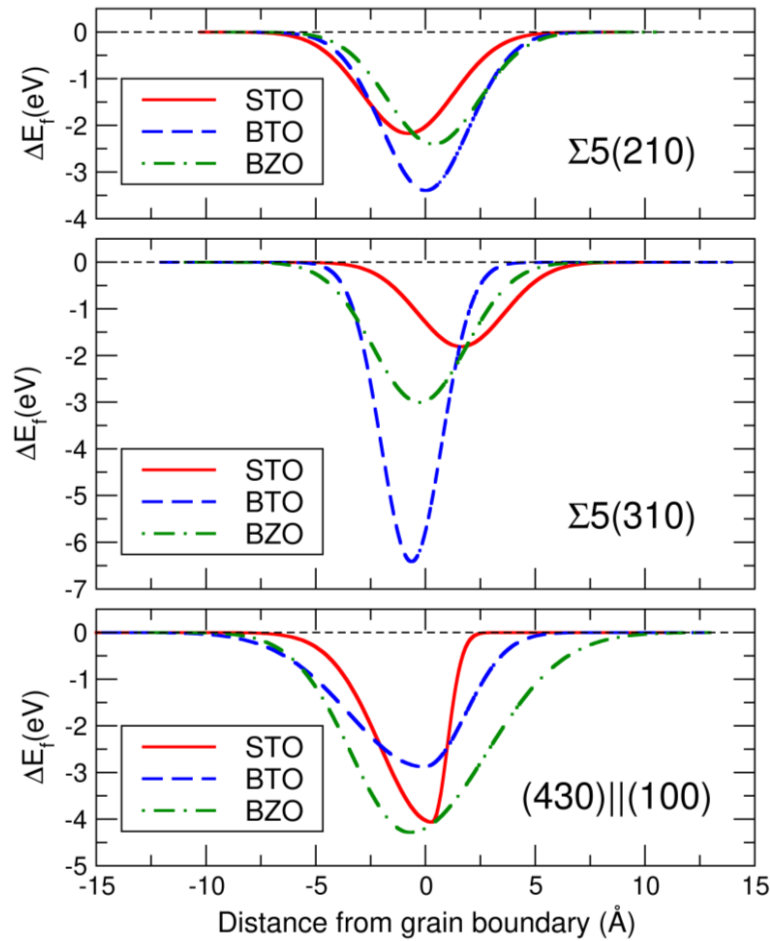

*Figure SI-8. Gaussian profiles [resulting from **method (2)**] of the negative oxygen vacancy formation energies with respect to values in the bulk grain interior for the three GB types in the three considered perovskite compounds.*